**ORIGINAL PAPER**

# Advancing the cybersecurity of the healthcare system with self-optimising and self-adaptative artificial intelligence (part 2)

Petar Radanliev[1] · David De Roure[1]

**Abstract**
This article advances the knowledge on teaching and training new artificial intelligence algorithms, for securing, preparing, and adapting the healthcare system to cope with future pandemics. The core objective is to develop a concept healthcare system supported by autonomous artificial intelligence that can use edge health devices with real-time data. The article constructs two case scenarios for applying cybersecurity with autonomous artificial intelligence for (1) self-optimising predictive cyber risk analytics of failures in healthcare systems during a Disease X event (i.e., undefined future pandemic), and (2) self-adaptive forecasting of medical production and supply chain bottlenecks during future pandemics. To construct the two testing scenarios, the article uses the case of Covid-19 to synthesise data for the algorithms – i.e., for optimising and securing digital healthcare systems in anticipation of Disease X. The testing scenarios are built to tackle the logistical challenges and disruption of complex production and supply chains for vaccine distribution with optimisation algorithms.

**Keywords** Healthcare systems · Self-optimising artificial intelligence · Self-adaptative artificial intelligence · Disease X · Covid-19

## 1 Introduction

Artificial Intelligence (AI) 'is gradually changing medical practice' with biomedical applications and medical AI systems grounded on 'data acquisition, machine learning and computing infrastructure' [1]. AI is already used in major disease areas e.g., cancer, neurology, and cardiology [2]. AI offers the *'transformative potential'* that is similar in significance to the industrial revolution [3]. One aspect of this transformation is *'the provision of healthcare services in resource-poor settings'*, and the specific interest is the promise of overcoming health system hurdles with *'with the use of AI and other complementary emerging technologies'* [4]. Another aspect of this transformation is the *'ambient intelligence… empowering people's capabilities by the means of digital environments that are sensitive, adaptive, and responsive to human needs, habits, gestures, and emotions'* [5]. In the future, AI will *'help mental health practitioners re-define mental illnesses more objectively'*, by identifying *'illnesses at an earlier or prodromal stage when interventions may be more effective, and personalize treatments based on an individual's unique characteristics'* [6]. The transformation empowering technologies include (a) communication systems, (b) sensing nodes, (c) processors, and (d) algorithms, but the use of AI has transformed healthcare systems by analysing data in the fog/edge [7].

There are also concerns on safety and dangers of AI, for example one study investigated the adoption of IBM Watson in public healthcare in China [8] and argued that AI is a 'black box' because we cannot see the learning process and we don't really know when the AI has some problem. The 'black box' operations of current AI algorithms 'have resulted in a lack of accountability and trust in the decisions made', hence, we need 'accountability. transparency, result tracing, and model improvement in the domain of healthcare' [9]. The balance between values and risks is explained as *'there are some improvements that benefit and there are some challenges that may harm'* [10] and the deployment of AI in healthcare is facing 'challenges from both an ethical and privacy standpoint' [11]. Despite the concerns, AI is already used for multi-disease prediction [12] and has already been used in combination with healthcare robotics,

✉ Petar Radanliev
petar.radanliev@eng.ox.ac.uk

1   Department of Engineering Sciences, University of Oxford, OX1 3QG Oxford, England, UK

genetics AI data-driven medicine, AI-powered Stethoscope [13] and *'technologies such as the Internet of Things (IoT), Unmanned Aerial Vehicles (UAVs), blockchain, and 5G, to help mitigate the impact of COVID-19 outbreak'* [14]. Conversely, the AI 'algorithms that feature prominently in research literature are in fact not, for the most part, executable at the frontlines of clinical practice' [15]. The main reasons behind this are that adding AI in a fragmented healthcare system won't produce the desired results. Healthcare organisations also *'lack the data infrastructure required to collect the data needed to optimally train algorithms'* [15].

## 1.1 Background of the study

This article builds upon the design for cyber risk assessment of healthcare systems operating with artificial Intelligence and Industry 4.0 [16] and postulates that it is possible to forecast cyber risk in digital health systems with predictive algorithms. This postulate is evaluated with two case study scenarios, constructed for testing, and improving the autonomous artificial intelligence (AutoAI) algorithm functions. The first case study is founded on a predictive (forecasting) solution for helping healthcare systems cope with unpredictable events (e.g., Disease X). The second case study is founded on forecasting production and supply chain bottlenecks in future global pandemics. The first case scenario is developed to test the algorithms with real-time intelligence (data) from healthcare edge devices. The second case scenario is constructed in an Industry 4.0 production and supply chain system, integrated into a dynamic and self-adapting automated forecasting engine.

Strong security and privacy protocols are followed in the testing scenarios. The two case studies test the performance of the AutoAI as algorithmic-edge computing solutions that will enable much greater speed in future responses, facilitated by the increased connectivity of humans and devices. Secondly, the case studies will develop a new self-adaptive version of AutoAI to compete with the performance of current deep learning algorithms. The second case is constructed in the concept of the factories of the future i.e., 'Industry 4.0' as the core case for improving the future vaccine production and supply chain capacity and speed. The two case studies are grounded on the need to accelerate new algorithms to operate on new supply chain edge technologies e.g., IoT, drones, autonomous vehicles, robots, and 3D printing. An additional motivation is the need to integrate cybersecurity in the vaccine cold chain/supply chain and assess the cyber risks from using modern edge technologies on complex healthcare systems – often operating with legacy IT systems. We can anticipate that vaccine supply chains for Disease X will face many bottlenecks. The overall goal of the article is to apply an incremental approach for testing and improving the new compact AutoAI algorithms and to work on resolving problems as they emerge. By using the algorithms and knowledge developed in existing literature, this article engages with the construction of the two scenarios with new and emerging forms of data (NEFD) and applies a new model of AutoAI into a real-world predictive (forecasting) scenario. The output of the two case study scenarios is a new forecasting engine that can improve the vaccine production and supply chain capacity and speed and enhance the capability of digital healthcare systems to adapt to a Disease X event.

## 2 Methodology - data collection

(1) The first phase of the research applies a quantitative 'causal-comparative design' on already synthesised (secondary) training data from cyber actions and events that already occurred. Then, by applying quantitative 'correlational research' the statistical relationship is assessed between the *primary* and *secondary* failures and measure the actual risk [17]. Finding data for scenario constructions is a real challenge, especially *secondary* data. This concern was addressed with *stratified sampling* and *random sampling* in the data analysis. This approach will enable testing of the model at different stages. (2) The new AutoAI algorithm will need to process large amounts of data to predict and forecast future events. This will require large training datasets to 'teach' the algorithm how to forecast. The process is slow and difficult to adapt to managing fast-changing events. To resolve this, the research uses open data from connected devices. By using anonymised data from existing and established platforms, the new AutoAI algorithm can be trained with real-time data, based on human and technology interactions at scale. This approach secures mass participation and large sample sizes. In addition, the training scenario constructions use standard open-source intelligence (OSINT) to gather public domain data, including public repositories (e.g., shodan.io).

## 3 Two case study scenarios

### 3.1 Case study a): applying the novel AutoAI to resolving the cyber risk problem

Cyber-risk is a growing concern in our society and the risks are becoming more sophisticated [18], exposing our healthcare systems and critical infrastructure. The first case study constructs scenarios for testing the new AutoAI algorithm for forecasting cyber-risks from the digitalisation of healthcare during pandemics. This generated an increased attack surface and presented a diverse set of uncertain and unpredictive attack vectors. The increased cyber-risk level will integrate cybersecurity experts, artificial intelligence, and big data scientists at the forefront of digital healthcare. Therefore, the Covid-19 pandemic created the ideal scenario for testing AutoAI algorithms. The first case study scenario is constructed for predicting (forecasting) the increased cyber risks surface [19] in digital healthcare systems operating at the edge.

Such a method was needed for risk assessing digital health systems [20] even before Covid-19, but the pandemic has placed enormous pressure on the health systems around the globe. To forecast cyber risks in digital healthcare at the edge, firstly the algorithms will be used to measure the cyber-attack frequency and severity. Secondly, for analysing real-time data and predicting cyber risks at the edge, the scenario will use statistical methods e.g., confidence intervals and time-bound ranges for testing and improving the novel AutoAI algorithm. Thirdly, the case scenario applied a dynamic methodology for testing the novel compact version of AutoAI algorithms, by deploying the algorithms on edge devices e.g., IoT, and drones. As more digital health devices are connected, the risk surface increases and the risk of cascading effect also increases. In the spirit of incremental research, the novel AutoAI algorithm can be adapted for use with the FAIR method[1]. The AutoAI is applied with the FAIR method for classifying data into *primary* (i.e., health system response to a cyber event) and *secondary* (i.e., the failure of other systems as the reactions to a cyber event). In the first case study scenario, the AutoAI and the FAIR method are integrated by using a Bayesian optimisation as a probabilistic iterative algorithm based on two components (1) a surrogate process e.g., a Gaussian process or a tree-based model; and (2) an acquisition function to balance the training data 'exploration' and 'exploitation'. This will advance the current state-of-the-art cyber-risk assessment, which is based on compliance-based qualitative methodologies (e.g., heath map/traffic lights systems). The first scenario for testing and verifying the novel AutoAI model in a real-world case study scenario improves the state-of-the-art in cyber risk assessment by intersecting previously isolated disciplines (e.g., IoT, health care; neuromorphic computing) with AI algorithms (e.g., ANN, CNNs, classification, regression) and analytical methodologies, (e.g., confidence intervals, time-bound ranges) with a wide range of source evidence (e.g., spatiotemporal data, high-dimensional data, time-stamped data).

## 3.2 Case study b): applying the novel AutoAI for resolving vaccine production and supply chain bottlenecks

The second case scenario is constructed for preparing the healthcare system for a Disease X event. The training set from the first case study is advanced and tested on training data from the Covid-19 pandemic. The current application of AI for managing Covid-19 will provide some of the secondary training data we need to construct the scenario. The rest of the training data is collected from OSINT data sources. This scenario helps improve the algorithm and benefits society because it forecast the points of failure and helps the healthcare systems to cope with future pandemics. Constructing the second case study scenario analyses how a disruptive Disease X event, combined with disruptive new edge systems, and novel AI-based technologies, challenges the benefit to the society from technological advancements.

These events in combination increase their impact and escalate the existing production and supply chain bottlenecks. By integrating AI in healthcare edge analytics, this case scenario devises a new approach for cognitive data analytics in vaccine production and supply chain. Creating a stronger resilience through cognition, resolving around understanding how and when bottlenecks and compromises happen, enabling healthcare systems to continuously adapt to global pandemics and employ AI techniques to understand and mitigate the cyber vulnerabilities of future adverse events i.e., Disease X. The second case scenario constructs and tests algorithmic solutions for protecting healthcare systems. The second case scenario is constructed for the AI algorithm to be tested on cloud computing.

## 4 Conceptual design

The conceptual design is grounded on the postulate that it is possible to construct predictive algorithms that can forecast risk in digital health systems operating at the edge of the network and severely improve the health care system's ability to rapidly adapt to huge external changes.

The first scenario is constructed to forecast cyber risk events during future pandemics, and the second is for testing the AI algorithms on digital healthcare devices operating on the edge of the network. These scenarios are designed to resolve two contemporary problems. Covid-19 has incentivised the adoption and scale-up of digital healthcare technologies. The first scenario forecast such cyber risk by creating urgency in predicting (forecasting) the potential risk of these complex and coupled systems. Although AI has been used in cybersecurity, the current AI algorithms cannot run on IoT devices with very low memory. The new

---

[1] https://www.fairinstitute.org/.

AutoAI algorithm makes this possible, through faster and more efficient processing. This scenario forecasts the points of failure from data collected on the edge of the network.

## 4.1 Design phases

Phase 1 synthesises the knowledge from existing studies to build the training scenarios for constructing a self-adaptive AI that can predict (forecast) the cyber risk in healthcare systems during a future Disease X event. Phase 2 extends into the development of a novel self-adaptive version of the AutoAI, that can severely improve the health care system's ability to rapidly adapt to huge external changes. The self-adaptive AI will not only secure the system by forecasting risk in Phase 1 but also address in Phase 2 how the system responds in circumstances when failure and compromise occur. This research acknowledges that not all systems can be secured and places the efforts on constructing self-adaptive algorithms. In healthcare systems, this is a high-risk strategy, because of the patient data confidentiality and the risk of compromised data resulting in a loss of life. However, in times of crisis, such as the crises we have seen during Covid-19, any disruption to the healthcare system could lead to loss of life. Since both options are not acceptable, as an alternative option is proposed to (apply the AutoAI algorithm to) develop a self-optimising and self-adaptive healthcare.

## 4.2 Phase 1: Predicting (forecasting) cyber risk in healthcare systems during a Disease X event

Phase 1 engages with designing a new AI training scenario for predicting (forecasting) the cyber risks emerging from increased digitalisation during global pandemics (e.g., working from home during lockdowns).

The first design obstacle $O_1$ is to automate the Bayesian approach with new AI algorithms and discrete binary (Bernuolli) probability distribution. The first design milestone ($M_1$) is to integrate and test if the novel AutoAI algorithm can operate on healthcare edge devices to detect (and forecast) anomalies before they turn into faults. One approach for this to be achieved is by applying apply discrete binary (Bernuolli) probability distribution. Alternatively, a second approach is to apply continuous probability distribution to determine the range and impact of a cyber-attack (size and duration) on edge devices. Both approaches will require automating the data preparation from edge devices. Such solutions are non-existent at present and require going beyond the current state-of-the-art to construct a more compact and efficient version of AI algorithms. The development of a new more compact and efficient versions of AI algorithms will create many additional benefits for healthcare and support services. The new self-adaptive AutoAI algorithms will be able to adapt to these changes because such AI will be capable of automated data collection, processing, and analysis of open-source intelligence (OSINT). The OSINT data accumulation in specific datacentres has created significant incentives and benefits for hackers to break into these datacentres. Hackers are already using deep learning methods to analyse failed attempts and to improve future attacks. The new self-adaptive, more compact, and efficient AutoAI will reduce the incentives for creating such data centres, by enabling the deployment of AI in edge devices (i.e., IoT devices, drones) for real-time automated data collection, processing, and analysis of raw data. The second design obstacle $O_2$ is to predict (forecast) risks and be more prepared for Disease X by progressing from manual to automated and from qualitative and quantitative risk assessment of failures in healthcare systems. The design methods include 'causal-comparative design' for synthesising data on actions and events that occurred during Covid-19 and 'correlational research' to assess the statistical relationship between the *primary* and *secondary* failures [17]. A *primary* failure will be classified as the failure of the health system caused by an event (e.g., Covid-19), and the *secondary* as the failure of other health systems as the reactions to an event (e.g., delayed surgeries, delayed cancer diagnosis, and/or treatment). The second design milestone ($M_2$) is to test if the novel AutoAI algorithm can forecast the actual loss including *primary* and *secondary* risk/loss from a Disease X event. This differentiates from forecasting risk/loss based only on *primary* failures. This is unusual because Disease X is an extraordinary event that could trigger catastrophic loss of life. The differences between primary and secondary loss are further described in recent literature [16]. In summary, the *primary* risk can easily be calculated and measured, but there are no mechanisms for reporting *secondary* risk. To address the (un)availability of data, the search for probabilistic data expands in new and emerging forms of OSINT data e.g., open data, spatiotemporal data, high-dimensional data, time-stamped data, and real-time data. Collecting and analysing such big data on a global scale is challenging, even with an AI-assisted approach for data collection, filtering, processing, and classification. To reduce complexity *stratified sampling* and *random sampling* will be applied to obtain reliable data samples. The third design obstacle $O_3$ is for the FAIR method[2] to be adapted and used for the data analysis and the comparative risk/loss aspects of the collected data. This leads to the third design milestone ($M_3$), advancing the integration of AI algorithms with established manual approaches for statistical risk assessment. Automation of the FAIR method will present reaching a significant milestone in advancement from manual to automated risk assessment. The unsupervised learning will be based on the postulate that failures will occur during a Disease X event. By forecasting failures, a solution can be devised to address how the system responds in these circumstances. In addition, by forecasting *primary* and *secondary* failures, the risk magnitude and the threat event frequencies can also be predicted. Automation of the FAIR method needs to combine risk analytics (e.g., FAIR), AI algorithms (e.g., ANN, CNNs, classification, regression), and statistical/analytical methodologies (e.g., confidence intervals; time bound ranges) with source evidence from a wide range of edge computing healthcare data, e.g., spatiotemporal data, high-dimensional data; time-stamped data. One of the main motivating points for intersecting the previously isolated disciplines is timing. Advancements in AI and edge computing present a new for studying global pandemics, and global pandemics such as Covid-19 are very rare events. Phase 1 of the $WP_3$ undertakes experimental developments in research on progressing from the current state of manual and qualitative risk assessment into an automated and quantitative risk analytical approaches. Alternatives to mitigate this risk of failing in $M_{1-3}$ include designing multiple highly specific AI algorithms for solving specific problems, then connecting the output. Although designing one transferable algorithm is preferred, this plan could be presenting lower risk and less complexity.

---
[2] https://www.fairinstitute.org/fair-u.

### 4.3 Phase 2: AI for vaccine development and self-adaptive production and supply chain bottlenecks

Phase 2 engages with designing and testing the AutoAI as an algorithmic solution for the medical production and supply chain bottlenecks (e.g., vaccines, protective equipment) during global pandemics i.e., Disease X. Although the Influenza vaccines are faster than ever to produce, the production and supply process still takes many months. To prepare the healthcare system for future pandemics, the healthcare system has two options, the first is to stockpile medications and the second is to improve the medical production and supply chain capacity and speed. Even if the vaccines and medications for future pandemics existed, stockpiling would be a challenge. Therefore, Phase 2 undertakes experimental developments in research on designing an adaptive algorithm for integrating the medical production and supply chains into the Industry 4.0 concept. Phase 2 is founded upon knowledge from Covid-19. The fourth design obstacle $O_4$ (continuing from Phase 1) is to use and test how the novel AutoAI algorithm can identify and adapt modern technologies. The fourth design milestone $M_4$ is for the novel AutoAI algorithm to be applied to resolve production and supply chain bottlenecks i.e., adapting to Industry 4.0 production and supply chains. While Phase 1 is using AI for cyber risk forecasting, Phase 2 is using AI for forecasting production and supply chain bottlenecks. The first bottleneck addressed is the need to secure the high-value medical products (e.g., vaccine) deliveries from theft, and sabotage. The second bottleneck addressed is the optimisation of the medical system resilience (i.e., the requirement to vaccinate the individuals operating the production/supply chain before starting a large-scale vaccination). Triggering optimisation challenges in terms of finding the fastest and safest method to use autonomous machines (e.g., drones) to deliver vaccines. The third bottleneck to address is the lack of coordination and risk from shortages. The fifth design milestone $M_5$ is to apply (and test) the novel AutoAI algorithm to operate on emerging new technologies (IoT, drones, autonomous vehicles, robots, 3D printing, etc.), and mixed design milestone $M_6$: is to integrate the AutoAI algorithm on complex healthcare systems (often operating with legacy IT systems) to ensure strong cybersecurity in the medical production and supply chains. The $WP_3$ requires a more technical approach for applying the new AutoAI in real-world scenarios for resolving practical problems e.g., cybersecurity, Disease X, production, and supply chain bottlenecks. The design obstacles and milestones are summarised in Table 4 below.

The expected difficulties in the conceptual design include the lack of suitable real-world medical testbeds (i.e., experimental technology). Numerous testbeds exist in controlled healthcare environments (e.g., EIT[3]), but the main characteristic of Disease X is unpredictability. Even if a real-world testbed can be secured in a Covid-19 scenario, the next pandemic might be defined with different characteristics. As an alternative to reaching the design milestones, the AutoAI can be tested in existing real-world production and supply chain testbeds designed for similar purposes (e.g., IIC[4]). Secondly, a small-scale low-cost autonomous supply chain testbed can be built. The second solution also enables testing the algorithms more extensively, with randomness built with Monte Carlo simulations to test for unexpected scenarios

**Table 4** Conceptual design

| M | Design obstacles (O), and design milestones (M) | O |
|---|---|---|
| **Phase 1**: Predicting (forecasting) cyber risk in healthcare systems during a Disease X event. | | |
| $M_1$ | Construct an automated Bayesian approach with the novel AutoAI algorithms and discrete binary (Bernoulli) probability distribution to forecast data breaches and the probability of a system going down. | $O_1$ |
| $M_2$ | Test if the novel AutoAI algorithm can forecast the **actual loss** including *primary* and *secondary* risk/loss from a Disease X event - with new and emerging forms of data. | $O_2$ |
| $M_3$ | Advance the FAIR method with unsupervised learning and regression algorithms and apply the AutoAI to forecast the cyber security readiness for Disease X event. | $O_3$ |
| **Phase 2**: Algorithmic solutions for medical production and supply chain bottlenecks. | | |
| $M_4$ | Test how the novel AutoAI algorithm can identify and adapt modern technologies to help with resolving shortages of supplies in critical times e.g., 3D printing. | $O_4$ |
| $M_5$ | Construct a self-adaptive vaccine delivery system based on the novel AutoAI algorithm, real-time data, and new modern technologies e.g., drones, autonomous vehicles, and robots. | $O_5$ |
| $M_6$ | Dynamic coordination with the novel AutoAI algorithmic design for real-time analytics of the vaccine supply chain in Disease X event. | $O_6$ |

## 5 Discussion on ethics and the gender dimension in AI

The emerging conceptual design can measure how AI infrastructures in the communications network and the relevant cybersecurity technology can evolve ethically that humans can understand while maintaining the maximum trust and privacy of the users. Particular attention is placed on the research methodology to eliminate gender and other types of bias in the participants' selection process for case study scenarios and interview participants. Participants (experts) are selected with established sampling techniques, based on expertise. A separate ethical concern included in the conceptual design is the issue of AI and gender equality. A report commissioned by UNESCO in 2019[5] found gender biases in AI training data sets, leading to algorithms and devices spreading and reinforcing harmful gender stereotypes and biases that are stigmatising and marginalising women on a global scale. Similarly, Google image recognition has been found to show bias toward African Americans[6]. The conceptual design investigates technical solutions for combating gender and racial bias with solutions such as *zeroing out* potential bias in words by setting some words to zero; and the use of less biased/more inclusive data such as multi-ethics, and multi-gender data. Such solutions could eliminate scenarios where people do not want to discriminate, but they do, though unintentional bias when the AI algorithms get biased data and make biased decisions. The proposed solutions can be applied in some obvious examples where data comes from unhealthy stereotypes. Some of the examples to be considered include a) gender stereotypes in healthcare systems, e.g., sex appeal - in which instances should AI be gender neutral? Male or female? Or racial bias – should AI be black or white?

## 6 Conclusion

The two case study scenarios present a practical application of a new research methodology for designing a self-optimising AutoAI capable of forecasting cyber risks in the health systems through real-time algorithmic analytics. The new methodology can be applied to designing a self-adaptive AutoAI specific for forecasting bottlenecks through autonomous analytics of digital health systems. Both scenarios use Covid-19 data to forecast cyber risks and bottlenecks in managing Disease X. This enhances the healthcare capacity in preparation for Disease X and create a comprehensive and systematic understanding of the opportunities and threats from migrating edge computing nodes AI technologies to the periphery of the internet. Enabling more active engagement of AI in the digitalisation of health systems, in response to the new IoT risk and security developments as they are emerging.

These concerns on IoT risk and security are fundamental to this work and the proposed solutions need to be applied with interdisciplinary research that works across engineering and computer science disciplines. The new methodology resolves a contemporary scientific problem that is relevant to users and the healthcare industry in general. The scenarios constructed in the article, assess the data protection and integrity while testing and improving the AI algorithms in edge devices. The new methodology contributes to knowledge by enabling the designing of compact and more efficient algorithms and combining statistics for securing the edge. Promoting secure developments in digital healthcare systems, through integrating AI algorithms in vaccine supply chains and cyber risk models.

### 6.1 Limitations and further research

The volume of data generated from edge devices creates diverse challenges in developing data strategies for training AI algorithms that can operate with lower memory requirements. Designing sparse compact and efficient AI algorithms for complex coupled healthcare systems demands prior data strategy optimisation and decision making on collecting and assessment of training data. In other words, the training data strategy should come before or simultaneously with the development of the algorithm. This is a particular risk concern because the new AI algorithm discussed in this proposal will lack such data input and must be tested in constructed scenarios. Hence, the new compact AI algorithm will be designed for a very specific function.

---

[3] https://eithealth.eu/catapult/.
[4] https://www.iiconsortium.org/vertical-markets/healthcare.htm.
[5] https://en.unesco.org/AI-and-GE-2020.
[6] https://www.cnbc.com/2018/09/18/world-economic-forum-ai-has-a-bias-problem-that-needs-to-be-fixed.html.

**Acknowledgements** Eternal gratitude to the Fulbright Visiting Scholar Project.

**Author Contributions (CRediT)**: Dr Petar Radanliev: Conceptualization, Data curation, Formal analysis, Investigation, Methodology, Visualization, Writing - original draft and review & editing. Prof. David De Roure: Funding acquisition, Methodology, Project administration, Resources, Software, Supervision, Validation, Writing—review & editing.

**Funding sources** This work was supported by the UK EPSRC [grant number: EP/S035362/1] and by the Cisco Research Centre [grant number CG1525381].

**Data Availability** all data and materials included in the article.

## Declarations

**Conflict of Interest** On behalf of all authors, the corresponding author states that there is no conflict of (or competing) interest.

## References

1. Yu KHsing, Beam AL, Kohane IS, "Artificial intelligence in healthcare," *Nat. Biomed. Eng.*, vol. 2, no. 10, pp. 719–731, Oct. 2018.
2. Jiang F, Jiang Y, Zhi H, Dong Y, Li H, Ma S, Wang Y, … Wang, Yongjun, "Artificial intelligence in healthcare: past, present and future," *Stroke Vasc. Neurol.*, vol. 2, no. 4, pp. 230–243, Dec. 2017.
3. Dwivedi YK, Hughes L, Ismagilova E, Aarts G, Coombs C, Crick T, Duan Y, … Williams MD. Artificial Intelligence (AI): Multidisciplinary perspectives on emerging challenges, opportunities, and agenda for research, practice and policy. Int J Inf Manage. Apr. 2021;57:101994.
4. Cossy-Gantner A, Germann S, Schwalbe NR, Wahl, Brian. Artificial intelligence (AI) and global health: how can AI contribute to health in resource-poor settings? BMJ Glob Heal. Aug. 2018;3(4):e000798.
5. Acampora G, Cook, Diane J, Rashidi P, Vasilakos AV, "A Survey on Ambient Intelligence in Health Care," *Proc. IEEE. Inst. Electr. Electron. Eng.*, vol. 101, no. 12, p. 2470, Dec. 2013.
6. Graham S, Depp C, Lee EE, Nebeker C, Tu X, Kim HCheol, Jeste DV, "Artificial Intelligence for Mental Health and Mental Illnesses: an Overview," *Curr. Psychiatry Rep.*, vol. 21, no. 11, Nov. 2019.
7. Qadri YAhmad, Nauman A, Zikria YBin, Vasilakos AV, Kim S, Won, "The Future of Healthcare Internet of Things: A Survey of Emerging Technologies," *IEEE Commun. Surv. Tutorials*, vol. 22, no. 2, pp. 1121–1167, Apr. 2020.
8. Sun TQian, Medaglia, Rony, "Mapping the challenges of Artificial Intelligence in the public sector: Evidence from public healthcare," *Gov. Inf. Q.*, vol. 36, no. 2, pp. 368–383, Apr. 2019.
9. Pawar U, O'Shea D, Rea S, O'Reilly R, "Explainable AI in Healthcare," *2020 Int. Conf. Cyber Situational Awareness, Data Anal. Assessment, Cyber SA 2020*, Jun. 2020.
10. Shaheen M, Yousef, "AI in Healthcare: medical and socio-economic benefits and challenges," *Sci. Prepr.*, Sep. 2021.
11. Bartoletti, Ivana. AI in healthcare: Ethical and privacy challenges. Lect Notes Comput Sci (including Subser Lect Notes Artif Intell Lect Notes Bioinformatics). 2019;11526 LNAI:7–10.
12. Xie S, Yu Z, Lv Z, "Multi-Disease Prediction Based on Deep Learning: A Survey," *Comput. Model. Eng. Sci.*, vol. 128, no. 2, p. 489, Jul. 2021.
13. Shaheen M, Yousef, "Applications of Artificial Intelligence (AI) in healthcare: A review," *Sci. Prepr.*, Sep. 2021.
14. Chamola V, Hassija V, Gupta V, Guizani, Mohsen. A Comprehensive Review of the COVID-19 Pandemic and the Role of IoT, Drones, AI, Blockchain, and 5G in Managing its Impact. IEEE Access. 2020;8:90225–65.
15. Panch T, Mattie H, Celi L, Anthony, "The 'inconvenient truth' about AI in healthcare," *npj Digit. Med.* 2019 21, vol. 2, no. 1, pp. 1–3, Aug. 2019.
16. Radanliev P, Roure D. David, "Disease X vaccine production and supply chains: Risk assessing healthcare systems operating with artificial Intelligence and Industry 4.0," 2022.
17. Eling M, Wirfs J, "What are the actual costs of cyber risk events?," *Eur. J. Oper. Res.*, vol. 272, no. 3, pp. 1109–1119, Feb. 2019.
18. Sardi A, Rizzi A, Sorano E, Guerrieri A. Cyber Risk in Health Facilities: A Systematic Literature Review. Sustainability. 2020;12(17):7002.
19. Ganin AA, Quach P, Panwar M, Collier ZA, Keisler JM, Marchese D, Linkov I, "Multicriteria Decision Framework for Cybersecurity Risk Assessment and Management," *Risk Anal.*, vol. 40, no. 1, pp. 183–199, Sep. 2017.
20. Islam SM, Riazul., Kwak D, Kabir MD, Humaun., Hossain M, Kwak K-S. The internet of things for health care: a comprehensive survey. IEEE access. 2015;3:678–708.